\documentclass[
superscriptaddress,
 amsmath,amssymb,
 aps,
prb,
floatfix,
]{revtex4}

\usepackage{graphicx}
\usepackage{dcolumn}
\usepackage{bm}

\begin{document}

\title{Induced effects by the substitution of Zn in Cu$_2$ZnSn$X_4$ ($X=$ S and Se)}

\author{Guohua Zhong}
\affiliation{Center for Photovoltaics and Solar Energy, Shenzhen Institutes of Advanced Technology, Chinese Academy of Sciences, Shenzhen, 518055, P. R. China}

\author{Kinfai Tse}
\affiliation{Department of Physics, The Chinese University of Hong Kong, Shatin, New Territory, Hong Kong, China}%

\author{Yiou Zhang}
\affiliation{Department of Physics, The Chinese University of Hong Kong, Shatin, New Territory, Hong Kong, China}%

\author{Xiaoguang Li}
\affiliation{Center for Photovoltaics and Solar Energy, Shenzhen Institutes of Advanced Technology, Chinese Academy of Sciences, Shenzhen, 518055, P. R. China}

\author{Li Huang}
\affiliation{Department of Physics, South University of Science and Technology of China, Shenzhen, 518055, P. R. China}

\author{Chunlei Yang}
\email{cl.yang@siat.ac.cn}
\affiliation{Center for Photovoltaics and Solar Energy, Shenzhen Institutes of Advanced Technology, Chinese Academy of Sciences, Shenzhen, 518055, P. R. China}

\author{Junyi Zhu}
\email{jyzhu@phy.cuhk.edu.hk}
\affiliation{Department of Physics, The Chinese University of Hong Kong, Shatin, New Territory, Hong Kong, China}%

\author{Zhi Zeng}
\affiliation{Key Laboratory of Materials Physics, Institute of Solid State Physics, Chinese Academy of Sciences, Hefei 230031, China}%

\author{Zhenyu Zhang}
\affiliation{International Center for Quantum Design of Functional Materials (ICQD), Hefei National Laboratory for Physical Sciences at Microscales, University of Science and Technology of China, Hefei, Anhui 230026, China}

\author{Xudong Xiao}
\affiliation{Center for Photovoltaics and Solar Energy, Shenzhen Institutes of Advanced Technology, Chinese Academy of Sciences, Shenzhen, 518055, P. R. China}
\affiliation{Department of Physics, The Chinese University of Hong Kong, Shatin, New Territory, Hong Kong, China}

\date{\today}

\begin{abstract}
Based on the density functional theory with hybrid functional approach, we have studied the structural and thermodynamic stabilities of Cu$_2M$Sn$X_4$ ($M=$ Zn, Mg, and Ca; $X=$ S and Se) alloy, and have further investigated the electronic and optical properties of stable Cu$_2$MgSnS$_4$ and Cu$_2$MgSnSe$_4$ phases. Thermal stability analysis indicates that Cu$_2$MgSnS$_4$ and Cu$_2$MgSnSe$_4$ are thermodynamically stable, while Cu$_2$CaSnS$_4$ and Cu$_2$CaSnSe$_4$ are unstable. The ground state configuration of the compound changes from kesterite into stannite structure when Zn atoms are substitued by larger Mg or Ca atoms. An energy separation between stannite and kesterite phase similar to that of CZTS is observed. Calculated electronic structures and optical properties suggest that Cu$_2$MgSnS$_4$ and Cu$_2$MgSnSe$_4$ can be efficient photovoltaic materials.

\end{abstract}

\maketitle
\section{Introduction}
Quaternary Cu$_{2}$ZnSnS$_{4}$ (CZTS) and Cu$_{2}$ZnSnSe$_{4}$ (CZTSe) compounds have drawn significant interests in the past years because of the abundance of their constituting elements, the environment-friendliness, and the appropriate band gaps ($1.4-1.6$ eV and $0.95-1.05$ eV for CZTS\cite{ref1,ref2,ref3,ref4,ref5,ref6} and CZTSe\cite{ref7,ref8,ref9}). They are regarded as a potential alternative to Cu(In,Ga)Se$_2$ (CIGSe), which is the most efficient thin-film solar cell material with a conversion efficiency record of 21.7\%.\cite{ref10} Mixing S and Se in Cu$_2$ZnSnS$_x$Se$_{4-x}$ (CZTSSe) can tune the band gap to achieve the best efficiency of 12.6\% so far.\cite{ref11} However, this efficiency is still lower than the requirement of commercially viable device. The physical origin of low efficiencies is complex. Gokmen \emph{et al.} pointed out that the existence of electrostatic potential fluctuation in the CZTS(Se) absorber layer could be an important factor that deteriorates the solar cell performances and the [Cu$_{\text{Zn}}$+Zn$_{\text{Cu}}$] defect cluster could be the origin for the electrostatic potential fluctuation.\cite{ref12} One trial to reduce the density of [Cu$_{\text{Zn}}$+Zn$_{\text{Cu}}$] defect cluster as proposed by Scragg \emph{et al.} is to design a thermal process with a low cooling rate after the high-temperature selenization or sulfurization process to obtain ordered CZTS.\cite{ref13} To eliminate the antisite defects originating from intermixing of Cu and Zn atoms which are actually with very similar atomic sizes, an alternative approach is to substitute other elements for Cu or Zn to possibly suppress the defects formation and thus improve the sample quality. In particular, the substitution large-size atoms for Cu or Zn atom is suggested in these Cu-poor CZTS and CZTSe.

For the substitution for Zn atom, Cd, Hg, Mg, Ca, and Sr as selected metals have been studied.\cite{ref14,ref15,ref16} These designs obviously increase the band gap of corresponding sulfides or selenides. From the stability analysis based on the generalized gradient approximation (GGA) calculation, the substitutions Cd and Hg for Zn are thermodynamic stable, while other substitutions are unstable against the phase-separation into the competing binary and ternary compounds. However, Cu$_2$MgGeS$_4$ nanoparticles have recently been synthesized, which shows X-ray diffraction pattern in good agreement with that of the theoretically proposed structure.\cite{ref17} Particularly, the synthesis of Cu$_2$MgSnS$_4$ nanoparticles without coexisting with binary compounds further implies the feasibility of Mg substitution for Zn element.\cite{ref18} For the contradiction, checking Wang's calculations, we conclude that GGA has difficulty in accurately describing the energy such as the band gap and Wang \emph{et al.}\cite{ref16} did not include all the related phases to analyze the thermodynamic stability. Additionally, both Cd and Hg are not environment-friendly. We, therefore in this work, focusing on large size ions Mg and Ca and using more precise method, are to investigate the induced effects by the substitution of Zn to explore the feasibility of Cu$_2M$Sn$X_4$ ($M=$ Mg and Ca; $X=$ S and Se) as photovoltaic materials.

\section{Computational Details}
The calculation is carried out within the framework of the density functional theory (DFT). The Vienna \emph{ab} \emph{initio} simulation package (VASP)\cite{ref19} is employed to optimize the crystal structures and calculate the stability and electronic and optical properties. The inner electrons are treated by the projector augmented wave method, while the valence electrons are expanded in plane waves with a cutoff energy of 400 eV. For the optimization, a conjugate-gradient algorithm is used to relax the ions into their instantaneous ground state. In all calculations, the Monkhorst-Pack \emph{k}-point grids are generated according to the specified \emph{k}-point separation 0.02 {\AA}$^{-1}$ and the convergence thresholds are set as $10^{-6}$ eV in energy and 0.005 eV/{\AA} in force.

In the standard DFT, the GGA of Perdew-Burke-Ernzerhof (PBE) version\cite{ref20} is adopted to describe the electronic exchange-correlation (XC) interactions. However, the standard DFT usually leads to erroneous descriptions for some real systems such as underestimating the band gap. One way of overcoming this deficiency is to use the Heyd-Scuseria-Ernzerhof (HSE) hybrid functional, where a part of the nonlocal Hartree-Fock (HF) type exchange is admixed with a semilocal XC functional, to give the following expression\cite{ref21}:
\begin{eqnarray}
E^{\text{HSE}}_{\text{xc}}=&&\alpha E^{\text{HF,SR}}_{\text{x}}(\nu)+(1-\alpha)E^{\text{PBE,SR}}_{\text{x}}(\nu)\nonumber \\
&&+E^{\text{PBE,LR}}_{\text{x}}(\nu)+E^{\text{PBE}}_{\text{c}},
\end{eqnarray}
where $\alpha$ is the mixing coefficient and $\nu$ is the screening parameter that controls the decomposition of the Coulomb kernel into short-range (SR) and long-range (LR) exchange contributions. In this calculation, the HSE exchange-correlation functional is set as 25\% mixing of screened HF exchange to PBE functional, namely, $\alpha=0.25$. Noticeably, all results presented in this paper are obtained from HSE calculations.

\section{Results and discussion}
\begin{table}[b]
\caption{\label{tab:table1}
Calculated total energies (meV per unit cell) relative to the most stable phase among KS, ST, and PMCA for Cu$_2M$Sn$X_4$ ($M=$ Zn, Mg, and Ca; $X=$ S and Se).}
\begin{ruledtabular}
\begin{tabular}{cccc}
system& $E_{\text{KS}}$ & $E_{\text{ST}}$ & $E_{\text{PMCA}}$ \\ \hline
CZTS  & 0        & 47.0     & 66.4       \\
CMTS  & 40.2     & 0        & 95.9       \\
CCTS  & 543.8    & 0        & 694.9      \\
CZTSe & 0        & 80.4     & 131.1      \\
CMTSe & 42.1     & 0        & 148.5      \\
CCTSe & 536.0    & 0        & 754.5      \\
\end{tabular}
\end{ruledtabular}
\end{table}
Quaternary I-II-IV-VI semiconductors derived from cation mutation of CuInSe$_2$-like I-III-VI compound may adopt one of the three structural phases (Fig. 1), i.e. kesterite structure (KS) with the space group of $I\bar{4}$, stannite structure (ST) with the space group of $I\bar{4}2m$, and the primitive mixed CuAu-like structure (PMCA) with the space group of $P\bar{4}2m$. While the first structure is derived from the chalcopyrite structure, the latter two are derived from the CuAu-like structure. The aforementioned structures were considered for the analysis of ground state structure of Cu$_2$MgSnS$_4$ (CMTS), Cu$_2$MgSnSe$_4$ (CMTSe), Cu$_2$CaSnS$_4$ (CCTS), and Cu$_2$CaSnSe$_4$ (CCTSe). From calculated total energies based on HSE functional (Table I), it is found that PMCA is the least stable structure among these three. While we have confirmed the previous calculation results showing that KS is the ground state structure for Zn-based compounds.\cite{ref22} On contrary, both Mg-based and Ca-based compounds are stabilized at the ST phase, which is similar to other substitution induced by large size atoms such as Cu$_2$CdSn$X_4$ and Cu$_2$HgSn$X_4$ ($X=$ S and Se). The energy difference between KS and ST is 47.0 meV per unit cell for CZTS and 80.4 meV per unit cell for CZTSe, which are according with Chen's results.\cite{ref15} In CMTS and CMTSe, this energy difference between KS and ST is 40.2 and 42.1 meV per unit cell, respectively. But this energy difference has a large of increase in CCTS and CCTSe. This value respectively reaches to 543.8 and 536.0 meV per unit cell for CCTS and CCTSe. The energy difference between ST and KS phase of CMTS and CMTSe is comparable to that of CZTS, hence CMTS and CMTSe based on ST phase are potentially stable at thermodynamic equilibrium.

As visualized from the crystal structure, the small total energy difference between the two structures could result in the existence of large amount of antisite defects. In Zn-based materials, based on the total energy difference value displayed in Table I, at high-temperature, the small total energy difference between the KS and PMCA phases in CZTS implies the easy formation of PMCA secondary phase or [Zn$_{\text{Sn}}$+Sn$_{\text{Zn}}$] related defect complex. In the Mg-based materials, the PMCA secondary phases can not easily form both in CMTS and CMTSe because of the relatively large total energy difference between the ST and PMCA structures. In Ca-based materials, the much larger atomic radius of Ca than that of Cu can induce a strong mechanical stress and can suppress the [Cu$_{\text{Ca}}$+Ca$_{\text{Cu}}$] complexes. The total energy difference between ST and PMCA also increases so that the PMCA secondary phases and [Ca$_{\text{Sn}}$+Sn$_{\text{Ca}}$] defect complexes can be suppressed.

\begin{table*}
\caption{\label{tab:table2} The optimized lattice constant, the anion displacement parameter $u$, the nearest atomic distance, and the calculated band gap for Cu$_2M$Sn$X_4$ ($M=$ Zn, Mg, and Ca; $X=$ S and Se)}
\begin{ruledtabular}
\begin{tabular}{cccccccccc}
          &            &            &                 &                   &                  &                   &\multicolumn{3}{c}{$E_g$ (eV)} \\
System    &$a$ ({\AA}) & $c$ ({\AA})& $u$             & $d_{\text{Cu}-X}$ ({\AA})& $d_{M-X}$ ({\AA})& $d_{\text{Sn}-X}$ ({\AA})& Expt.        & GGA  & HSE \\ \hline
CZTS (KS) & 5.475      & 10.943     & 0.25$\pm$0.0157 & 2.321 (2.325)     & 2.370            & 2.471             & 1.4$-$1.6\footnotemark[1]  & 0.09 & 1.60 \\
CMTS (ST) & 5.568      & 10.932     & 0.25$\pm$0.0006 & 2.330             & 2.471            & 2.464             & 1.63\footnotemark[2]       & 0.19 & 1.80 \\
CCTS (ST) & 5.903      & 10.483     & 0.25$\pm$0.0206 & 2.363             & 2.726            & 2.449             & $--$                       & 0.55 & 2.13 \\
CZTSe (KS)& 5.771      & 11.537     & 0.25$\pm$0.0183 & 2.437 (2.444)     & 2.499            & 2.623             & 0.95$-$1.05\footnotemark[3]& 0.02 & 1.00 \\
CMTSe (ST)& 5.864      & 11.562     & 0.25$\pm$0.0006 & 2.447             & 2.611            & 2.617             & $--$                       &-0.05 & 1.16 \\
CCTSe (ST)& 6.173      & 11.186     & 0.25$\pm$0.0185 & 2.476             & 2.859            & 2.602             & $--$                       & 0.18 & 1.62 \\
\end{tabular}
\end{ruledtabular}
\footnotetext[1]{Ref.~\onlinecite{ref1,ref2,ref3,ref4,ref5,ref6}.}
\footnotetext[2]{Ref.~\onlinecite{ref18}.}
\footnotetext[3]{Ref.~\onlinecite{ref7,ref8,ref9}.}
\end{table*}
For the ground-state structure of Cu$_2M$Sn$X_4$ ($M=$ Zn, Mg, and Ca; $X=$ S and Se), the lattice constants are listed in Table II to compare the substitution effects. The substitution of Mg for Zn leads to expansions along $a$ and $b$ directions. From the atomic distances, the bond lengths of $d_{\text{Cu}-X}$ and $d_{M-X}$ is found to increase with $M$ atomic radius. Except for CMTSe, the substitution induces a shrink along $c$ direction, which is possibly related to the shortening of bond length of $d_{\text{Sn}-X}$ in Mg- and Ca-based compounds. Noticeably, the anion (S or Se) displacement parameter $u$ deviates from the ideal value of 0.25 for Zn- and Ca-based materials, while it is almost ideal ($u=0.25$) for Mg-based compounds.

At experiment, more stable secondary and ternary impurity phases often decrease the crystallizing quality of quaternary compounds, or even no stable quaternary phase is obtained. So we have also examined the thermodynamic stability of the materials above in order to provide a guidance for their synthesis. The requirement that the Cu$_2M$Sn$X_4$ ($M=$ Zn, Mg, and Ca; $X=$ S and Se) phase is more stable than any secondary phases at thermodynamic equilibrium imposes a restriction on the range of chemical potentials. Satisfying the constraint requires that
\begin{eqnarray}
&\Delta H_\text{f}(\text{Cu}_2M\text{Sn}X_4)\nonumber \\
&=2\Delta \mu_{\text{Cu}}+\Delta \mu_{M}+\Delta \mu_{\text{Sn}}+4\Delta \mu_{X},
\end{eqnarray}
and for any secondary phases
\begin{equation}
\Delta H_\text{f}(\text{secondary~phases}) > \sum_in_i\Delta\mu_i,
\end{equation}
where $\Delta H_\text{f}$ denotes the formation enthalpy of the corresponding phase, $n_i$ and $\Delta\mu_i$ denote the number of atoms and the chemical potential of the corresponding element $i$\cite{ref23}. Constrained by Eq. (2), only 3 out of 4 chemical potentials are independent. We hence choose by convention the chemical potentials of metallic elements to be independent variables. It is clear that any $\Delta\mu$ should have a value less than 0 to avoid segregation of elements. Binary and ternary compounds further restrict the range of stable chemical potential. In the calculation, element Cu, $M=$Zn/Mg/Ca, Sn and $X=$S/Se, binary metal sulfides and selenides Cu$_2X$, Cu$X$, $MX$, Sn$X$, Sn$X_2$, and ternary compounds Cu$_2$Sn$X_3$\cite{ref23}, Cu$_3$Sn$X_4$\cite{ref24} are the secondary phases that have been considered.

\begin{table*}
\caption{\label{tab:table3}
Listing of copper chemical potential limit and volume of stable chemical potential range for Cu$_2M$SnS$_4$ and Cu$_2M$SnSe$_4$ ($M=$ Zn, Mg, and Ca) showing that replacement of Zn by Mg forms energetically stable compound.}
\begin{ruledtabular}
\begin{tabular}{cccc}
                                                      & $M=$ Zn & $M=$ Mg    & $M=$ Ca \\ \hline
Quaternary sulfide $\mu_{\text{Cu}}$ range (eV)       & -0.72/0 & -0.67/0    & Unstable \\
Quaternary selenide $\mu_{\text{Cu}}$ range (eV)      & -0.65/0 & -0.62/-0.01& Unstable \\
Volume of stable chemical potential of Cu$_2M$SnS$_4$ & 0.076   & 0.038      & $--$ \\
Volume of stable chemical potential of Cu$_2M$SnSe$_4$& 0.057   & 0.023      & $--$ \\
\end{tabular}
\end{ruledtabular}
\end{table*}
We adopt the range of copper chemical potential and the volume of stable chemical potential as an indicator of the feasibility of growing the Cu$_2M$Sn$X_4$ material as copper vacancy is likely a candidate of $p$-type conductivity. The ternary phase Cu$_3$Sn$X_4$ is not involved in the boundary of stable chemical potential range when HSE exchange-correlation function is used. The copper chemical potential limits and volume of allowed chemical potential range are shown in Table III. The copper limit of Mg-based material is slightly smaller than that of Zn-based counterpart, with the volume of allowed chemical potential remaining 40$-$50\% of Zn-based compound, and Ca-based material is unstable with respect to its secondary phases. The mixing of Cu$_2$Sn$X_3$ and Mg$X$ to form Cu$_2$MgSn$X_4$ is thermodynamically favorable if
\begin{equation}
\Delta H_\text{f}(\text{Cu}_2\text{MgSn}X_4)\leq\Delta H_\text{f}(\text{Mg}X)+\Delta H_\text{f}(\text{Cu}_2\text{Sn}X_3)
\end{equation}
even with entropy effect not considered. The reason that previous work\cite{ref16} using GGA functional had not reveal the thermodynamic stable range of Cu$_2$MgSn$X_4$ is mainly due to GGA¡¯s inability to correctly evaluate ground state energy of Cu$_2$MgSnS$_4$ and Cu$_2$SnS$_3$. For example, a comparison of GGA result to that of HSE shows 0.98 eV rise in formation energy in Cu$_2$MgSnS$_4$, while the rise of formation energy for Cu$_2$SnS$_3$ and MgS is 0.62 eV and 0.25 eV. This subtle energy difference actually resulted in a significant change in the phase diagram as it completely eliminates the stable chemical potential region. The relative smaller volume of allowed chemical potential for Mg-based compound is likely due to the relatively larger internal stress of the material. Since ZnS (ZnSe), Cu$_2$SnS$_3$ (Cu$_2$SnSe$_3$) and CZTS (CZTSe) have similar lattice constants, the bond length between Zn and S (Se) in CZTS(Se) is almost identical to that in the binary ZnS (ZnSe) compounds. As a result, internal stress around Zn elements is expected to be small in these compounds. In contrast, the ionicity of Mg and Ca results in an energetically more stable halite-structured MgS(Se) and CaS(Se), with significantly longer bond length and higher coordination number than the zinc-blende counterpart. Atoms in Mg-based compounds are likely to be more stressed. This internal stress may lead to a relatively narrower stable chemical potential range for CMTS and CMTSe.

To evaluate the potential of CMTS and CMTSe as photovoltaic materials, however, we have further investigate their electronic and optical properties and compare them with that of Zn-based materials. As shown in Table II, GGA strongly underestimates the band gap. Adopting HSE hybrid functional, we reproduce the experimental band gaps of CZTS and CZTSe\cite{ref1,ref2,ref3,ref4,ref5,ref6,ref7,ref8,ref9}. ST structured CMTS and CMTSe are direct band gap materials, with calculated band gap as 1.80 eV and 1.16 eV respectively, which are correspondingly larger than those of CZTS and CZTSe. Ca-based materials have bigger band gap of 2.13 eV for CCTS and 1.62 eV for CCTSe, which mainly comes from the stronger hybridization interaction between Sn-$s$ and $X$-$p$ states as well as from the weaker coupling interaction between $M-d$ and $X-p$ states. Seen from the band structures and the projected density of states (PDOS) shown in Fig. 3, Mg substitution does not result in a change in nature of valence and conduction band and makes no contribution to the electronic states near the Fermi level. The valence bands mainly come from the hybridization of Cu$-3d$ and S$-3p$ (Se$-4p$) electronic states and the first-group conduction bands mainly come from the mixture of Sn$-5s$ and S$-3p$ (Se$-4p$) electronic states. The small increase of band gap in Mg-based materials mainly results from the enhancement of interaction between Sn and S (Se) induced by the bond length shortening comparing to Zn-based materials.

The optical properties of CMTS and CMTSe are compared with those of Zn-based materials. Here, the calculation of the optical properties is based on the complex dielectric function $\varepsilon(\omega)=\varepsilon_1(\omega)+i\varepsilon_2(\omega)$ describing the polarization response of the material to an externally applied electric field\cite{ref25}. The dielectric function of CMTS and CMTSe are shown in Fig. 4 and correspondingly compared with CZTS and CZTSe. The onset to the response of the imaginary part $\varepsilon_2(\omega)$ of the dielectric function reflects the band-gap energy, and in the vicinity of this onset of the ST CMTS and CMTSe show strong anisotropy $\varepsilon^{\text{zz}}_{2}>\varepsilon^{\text{xx}}_{2}$ compared to the KS CZTS and CZTSe. This is a consequence of the stronger and positive crystal-field splitting in the ST compounds. At somewhat larger energies (in the regions about $2.0-4.0$ eV for sulfides, $1.5-3.5$ eV for selenides) opposite anisotropy appears: $\varepsilon^{\text{xx}}_{2}(\omega)< \varepsilon^{\text{zz}}_{2}(\omega)$ for ST CMTS and CMTSe, and $\varepsilon^{\text{xx}}_{2}(\omega)> \varepsilon^{\text{zz}}_{2}(\omega)$ for KS CZTS and CZTSe. The anisotropy is reflected also in the real part $\varepsilon_1(\omega)$ of the dielectric function. The selenides show larger high-frequency dielectric constant $\varepsilon_{\infty}$ in accordance with the smaller band gap E$_g$. The ST Mg-based compounds show slightly smaller high-frequency dielectric constant than the KS Zn-based materials. The value of $\varepsilon_{\infty}$ is estimated to be about 3.9 and 4.8 for CMTS and CMTSe, respectively.

The absorption coefficient $\alpha(\omega)$ for ST CMTS and CMTSe, together with those for CZTS and CZTSe, for the optical absorption from the valence bands to the lowest conduction bands are shown in Fig. 5. Since the absorption is obtained directly from the dielectric function\cite{ref25}, similarities in the polarization response are reflected also in the absorption coefficient. Thus, all four compounds have comparable absorption, although with different photon energy for the onset. A large band-edge absorption coefficient of $\sim 10^5$ cm$^{-1}$ is observed for all the four materials.

\section{Conclusions}
Using the first-principles calculations and combining with the HSE hybrid functional, we have studied the induced effects by the substitution of Zn in Cu$_2$ZnSn$X_4$ ($X=$ S and Se), from structural stability, thermodynamic stability, electronic characteristics, and optical properties. Large size atom Mg and Ca substituting for Zn, we found that the ground-state structure is stabilized at ST configuration instead of KS as Zn-based compounds. The result of total energy difference between KS and ST or between ST and PMCA indicates that the substitution larger size Ca atom for Zn atom can efficiently suppress the antisite defects. Unfortunately, Ca-based materials are not thermodynamically stable with respect to the secondary phases. Mg-based materials are thermodynamic stable, with energy difference between the ST and KS phases comparable to that of CZTS and CZTSe. Additionally, [Mg$_{\text{Sn}}$+Sn$_{\text{Mg}}$] defect complexes are expected to be largely reduced in CMTS and CMTSe. Mg substitution does not change the electronic characteristics but opens a slightly bigger band gap than Zn-based compounds. The optical absorption spectra show the feasibility of Cu$_2$MgSnS$_4$ and Cu$_2$MgSnSe$_4$ as a photovoltaic material.

\begin{acknowledgments}
The work was supported by the National Major Science Research Program of China under Grant no. 2012CB933700, the Natural Science Foundation of China (Grant nos. 61274093, 61574157, 11274335, 11504398, 51302303, 51474132, and 11404160), and the Basic Research Program of Shenzhen (Grant nos. JCYJ20150529143500956, JCYJ20140901003939002, JCYJ20150521144320993, JCYJ20140417113430725, and JCYJ20150401145529035). XDX also wish to acknowledge the support by the Theme-based Research Scheme No. T23-407/13-N of Hong Kong Research Grant Council.
\end{acknowledgments}


\begin{figure}
\includegraphics[width=\columnwidth]{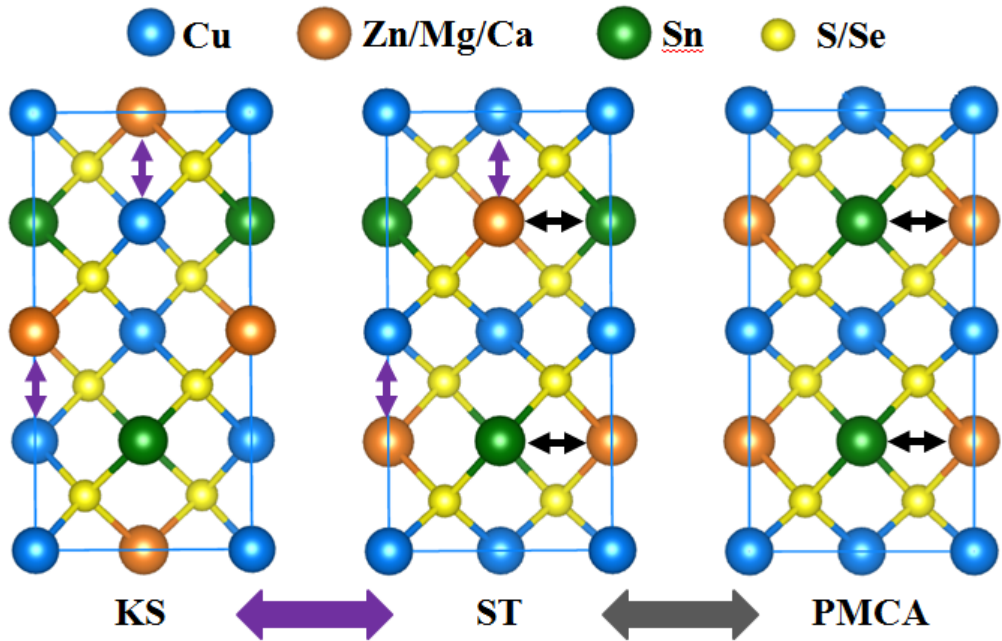}
\caption{(Color online) Three possible configurations of Cu$_2M$Sn$X_4$ ($M=$ Zn, Mg, and Ca; $X=$ S and Se): KS, ST, and PMCA. The purple arrows mark the [Cu$_{\text{Zn}}$+Zn$_{\text{Cu}}$] defects complexes that result in the mutual transformation between KS and ST structures. The black arrows mark the [Zn$_{\text{Sn}}$+Sn$_{\text{Zn}}$] defects complexes that result in the mutual transformation between ST and PMCA structures.}
\end{figure}

\begin{figure*}
\includegraphics[width=\textwidth]{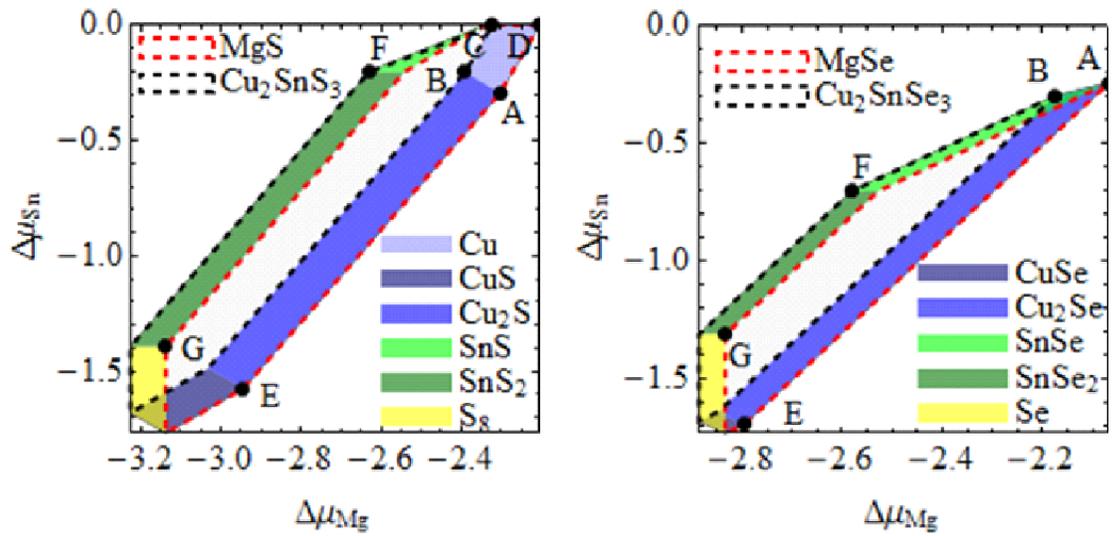}
\caption{(Color online) Illustration of phase diagram of CMTS and CMTSe. The large surface bound by Cu$_2$SnS$_3$ (Cu$_2$SnSe$_3$) is the front surface and that of MgS (MgSe) is the back surface. Indicators of Cu-chemical potential are plotted at Cu-richest limit (A-D), Cu-mild range (E-F) and Cu-poorest limit (G).}
\end{figure*}

\begin{figure*}
\includegraphics[width=\textwidth]{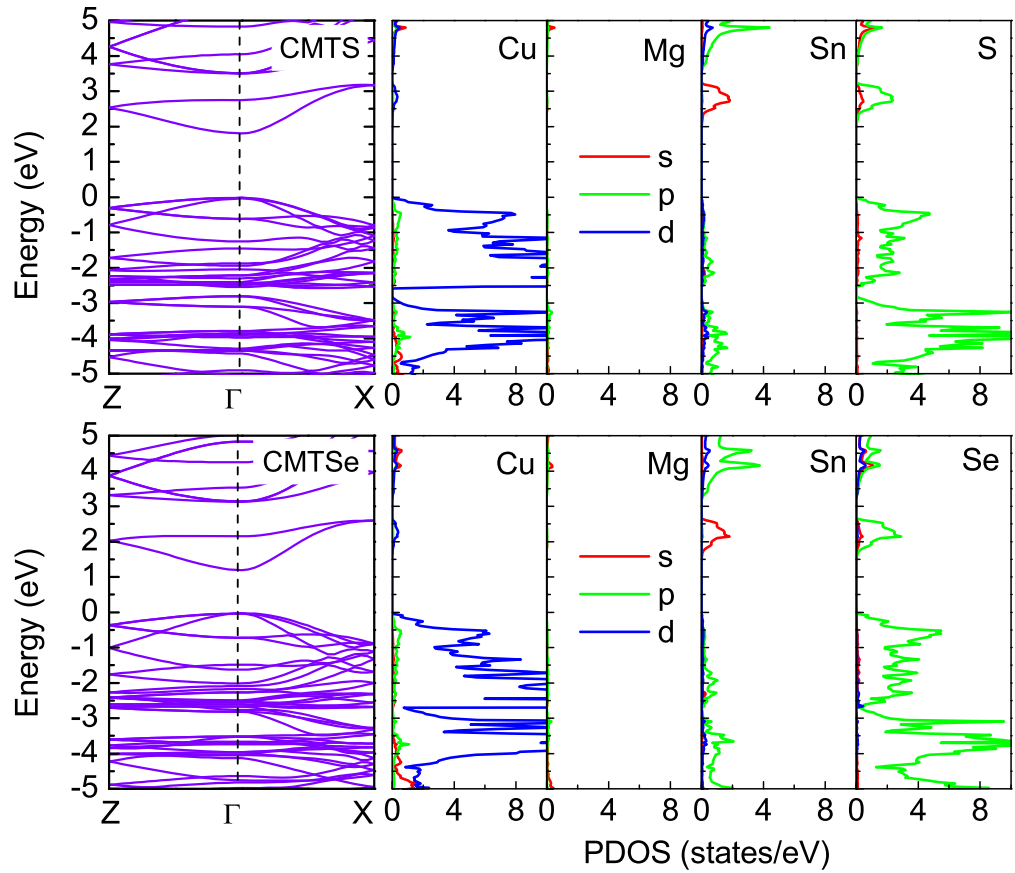}
\caption{(Color online) Calculated band structures and projected density of states (PDOS) for ST Cu$_2$MgSnS$_4$ (top) and Cu$_2$MgSnSe$_4$ (bottom).}
\end{figure*}

\begin{figure*}
\includegraphics[width=\textwidth]{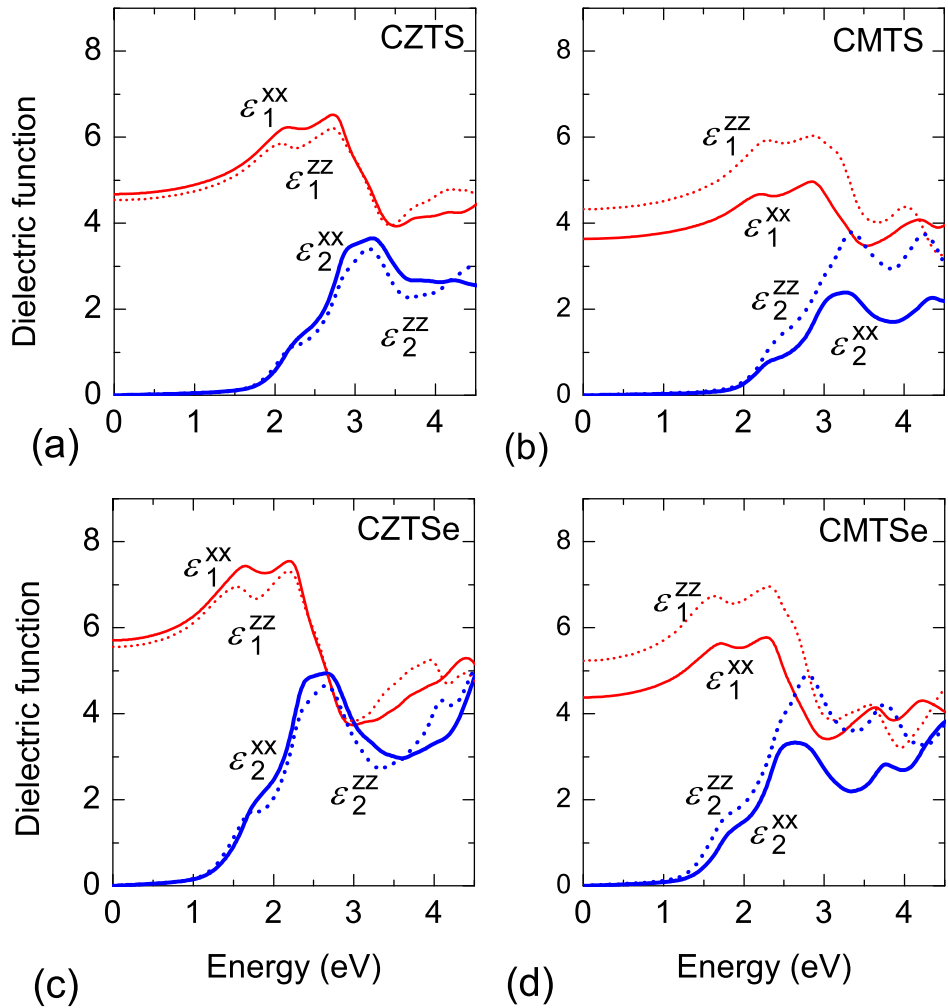}
\caption{(Color online) The dielectric function $\varepsilon(\omega)=\varepsilon_1(\omega)+i\varepsilon_2(\omega)$ of Cu$_2$MgSnS$_4$ and Cu$_2$MgSnSe$_4$. The red line represents the real part $\varepsilon_1(\omega)$ and the blue line represents the imaginary part $\varepsilon_2(\omega)$. The dielectric function is divided into the transverse contribution (xx; solid lines) and longitudinal contribution (zz; dotted lines). For comparison, the dielectric function of KS Cu$_2$ZnSnS$_4$ and Cu$_2$ZnSnSe$_4$ are also presented.}
\end{figure*}

\begin{figure}
\includegraphics[width=\columnwidth]{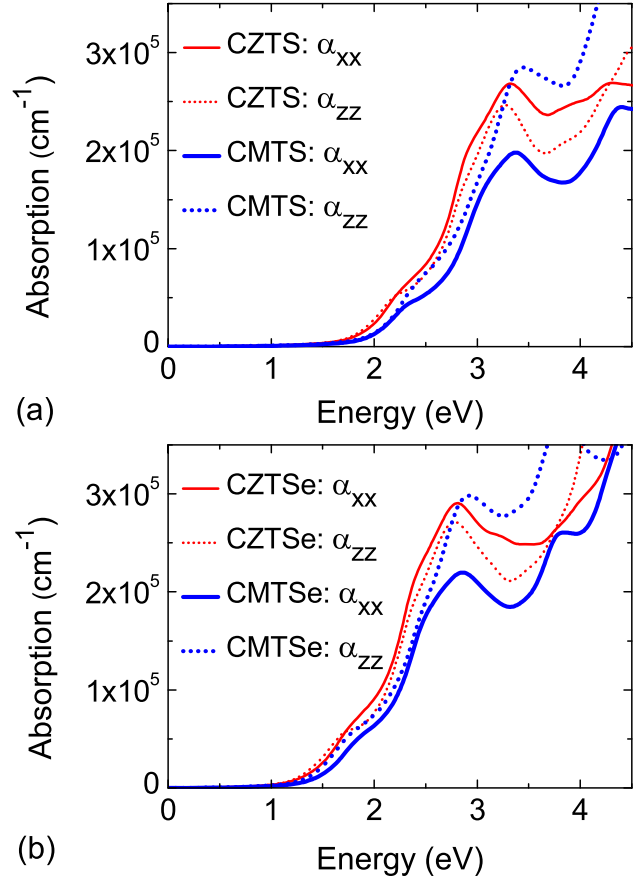}
\caption{(Color online) Calculated absorption coefficient $\alpha(\omega)$ in visible light region of ST Cu$_2$MgSnS$_4$ (a) and Cu$_2$MgSnSe$_4$ (b). For comparison, the absorption coefficient $\alpha(\omega)$ of KS Cu$_2$ZnSnS$_4$ and Cu$_2$ZnSnSe$_4$ is also presented.}
\end{figure}

\end{document}